\documentclass[aps,superscriptaddress,twocolumn,prl,showpacs]{revtex4-1}

\usepackage[centertags]{amsmath}
\usepackage{amsmath}
\usepackage{amssymb}
\usepackage{phaistos}
\usepackage{epsfig}
\usepackage{color}
\usepackage{amsthm}
\usepackage{graphicx}
\usepackage{float}
\usepackage{natbib}

\begin{document}
\title{Trapping cold ground state argon atoms for sympathetic cooling of molecules}
\author{P. D. Edmunds and P. F. Barker}
\affiliation{Department of Physics and Astronomy, University College London}

\begin{abstract}
We trap cold, ground-state, argon atoms in a deep optical dipole trap produced by a build-up cavity.  The atoms, which are a general source for the sympathetic cooling of molecules, are loaded in the trap by quenching them from a cloud of laser-cooled metastable argon atoms. Although the ground state atoms cannot be directly probed, we detect them by observing the collisional loss of co-trapped metastable argon atoms using a new type of  parametric loss spectroscopy.  Using this technique we also determine the polarizability ratio between the ground and the metastable 4s[3/2]$_2$ state to be 40$\pm6$ and find a polarisability of (7.3$\pm$1.1) $\times$10$^{-39}$ Cm$^2/$V for the metastable state. Finally, Penning and associative losses of metastable atoms, in the absence of light assisted collisions, are determined to be $(3.3\pm 0.8) \times 10^{-10}$ cm$^3$s$^{-1}$. 
\end{abstract}

\maketitle

The development of methods to create, control and manipulate the motion of cold complex molecules has, over the last ten years, allowed the study of atomic and molecular interactions with unprecedented precision. Cold molecules offer a new testbed for precision measurement \cite{loh2013precision} and the exploration of cold collisions and chemistry \cite{henson2012observation}. Cold polar molecules are seen as promising candidates for studying condensed matter physics \cite{micheli2006toolbox} and even quantum information science \cite{andre2006coherent}.  Of central importance to these applications has been the development of techniques to create translationally cold molecules that are either in their absolute internal ground state or a well defined internal ro-vibrational state.  However, while many slow complex species can now be produced by methods such as Stark and Zeeman deceleration \cite{vanhaecke2007multistage,narevicius2008stopping,bethlem1999decelerating}, these non-dissipative methods are realistically limited to temperatures in mK range since the high energy particles must be discarded to reduce the translational energy spread.  Truly dissipative methods are therefore required to further cool molecules into the sub-mK regime where quantum effects will become dominant \cite{shuman2010laser}.

Sympathetic cooling is a promising general method for dissipative cooling, but typical laser cooled species are generally reactive and cannot generally be utilised \cite{lara2006ultracold}.  Trapped noble gas atoms in their ground state appear to be an ideal candidate for the sympathetic cooling of molecules \cite{barker,barletta2009towards,mcnamara2006degenerate} as they are chemically inert and can be laser-cooled to $\mu$K temperatures in an excited metastable state.  Cold helium gas has already been used extensively to buffer gas cool many species but temperatures are limited to the 100 mK range \cite{weinstein1998magnetic}. In addition, as these atoms are in their absolute ground state, inelastic state changing collisions which can prevent efficient collisional cooling can be reduced or avoided. All of the noble gas atoms have been laser cooled in a metastable state \cite{stas2006homonuclear,kuppens2002approaching,tychkov2006metastable,spoden2005collisional,PhysRevA.73.023406,katori1994laser,walhout1995optical} and all but helium can be quenched to its non-reactive ground state by dipole allowed transitions. However, once in their ground state they have no magnetic moment and cannot be trapped in a magnetic trap. Finally, ground state noble gases are difficult to detect using CW laser spectroscopic methods because the first dipole allowed transitions are in the vacuum ultraviolet where no available CW laser sources exist. 

Ground state noble gas atoms can be trapped in an optical dipole trap, despite their ground state static polarizabilities being at least an order of magnitude smaller than typical laser cooled species, since large optical fields detuned far from resonance can be used to trap them. Such large CW fields can be produced in optical buildup cavities, which have previously been used to create deep traps for a range of atomic species \cite{Lee,Mosk,Cruz,eichhorn}. Additionally, species that cannot be directly observed can be detected when simultaneously trapped with another species that can be probed spectroscopically. This can be accomplished because the interactions between the two species in a trap perturb the motion of the observable species via intra-trap collisions. Examples include atomic ions that are sympathetically cooled by other trapped ions or ions produced by chemical reactions with others in the trap \cite{blythe2005production}. Detection is accomplished by modulating the trap potential to parametrically heat the species that cannot be directly observed. This frequency is usually unique to each trapped species because of their differences in mass. The modulation heats the species which can be observed and is detected as a change in the fluorescence monitored from the trap. By recording the trap fluorescence as a function of modulation frequency a type of species-specific mass spectrometry has been achieved in ion traps. In neutral atom traps parametric heating is a well established way of characterising the trap. For example, the loss induced by parametric heating is commonly used to identify trap frequency and therefore trap depth for a particular species \cite{vuletic1998degenerate}. In addition, by tuning slightly away from the parametric resonance, selective removal of hot atoms in the trap has been demonstrated \cite{poli2002cooling}.

\begin{figure}
\centering
\includegraphics[width=1\columnwidth]{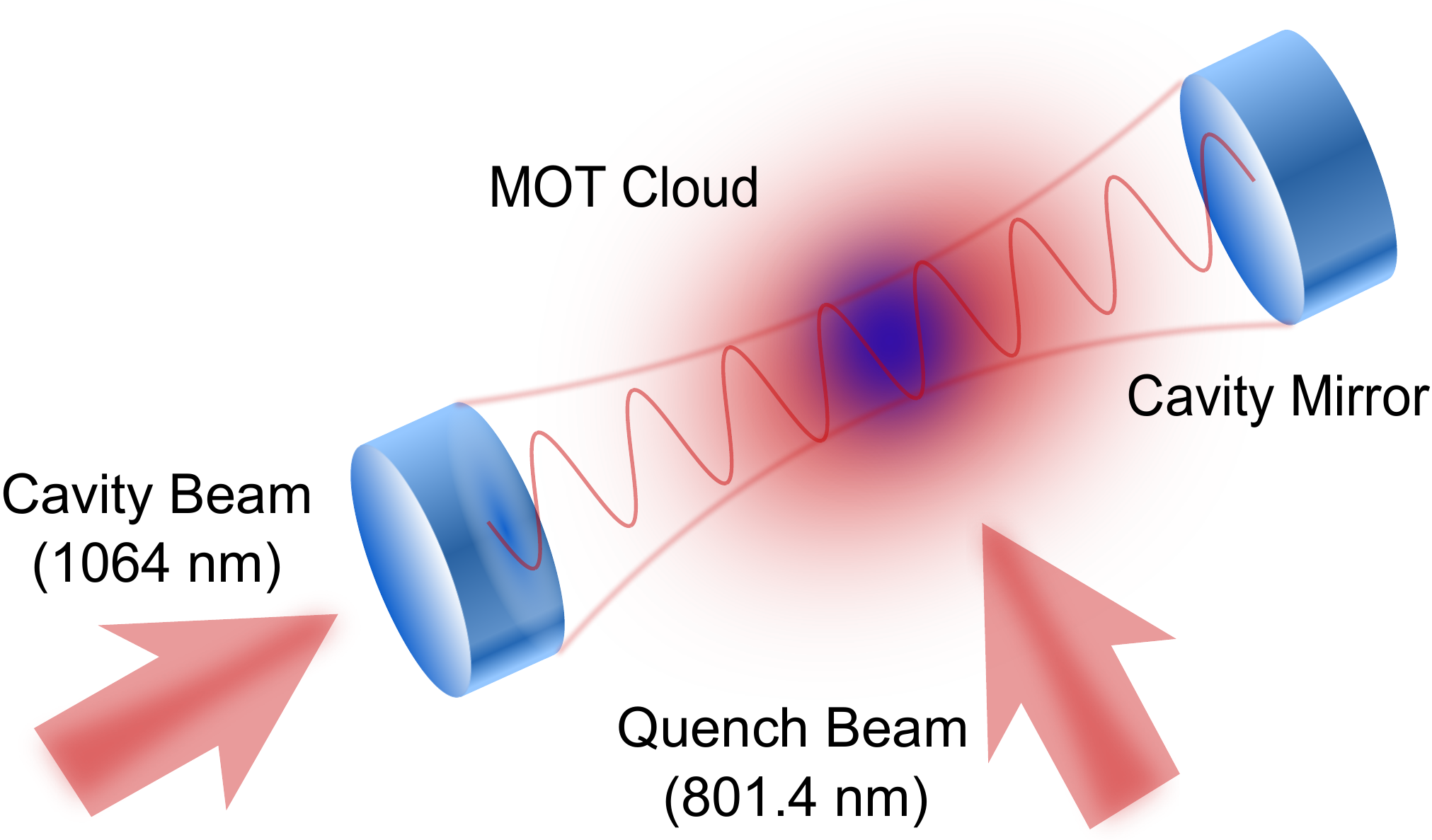}
\caption{Schematic of the optical cavity. Metastable argon is first cooled in a MOT, and then quenched down to the ground state. Both species can be trapped in the lattice formed within the optical build-up cavity.}
\label{fig:cavity}
\end{figure}

In this letter we describe dipole trapping of cold ground state argon atoms suitable for the sympathetic cooling of molecules.  We also demonstrate the detection of the ground state by using a type of parametric loss spectroscopy based on co-trapping a small fraction of metastable argon atoms. This allows us to simultaneously measure the presence of ground state argon atoms as well as the polarizability of the metastable state at the trapping wavelength of 1064 nm.  Finally, we measure for the first time, Penning and associative losses of trapped metastable atoms in the absence of resonant laser light and determine the loss rate for metastable atoms by collisions with the co-trapped ground state atoms. 

\begin{figure}
\centering
\includegraphics[width=0.7\columnwidth]{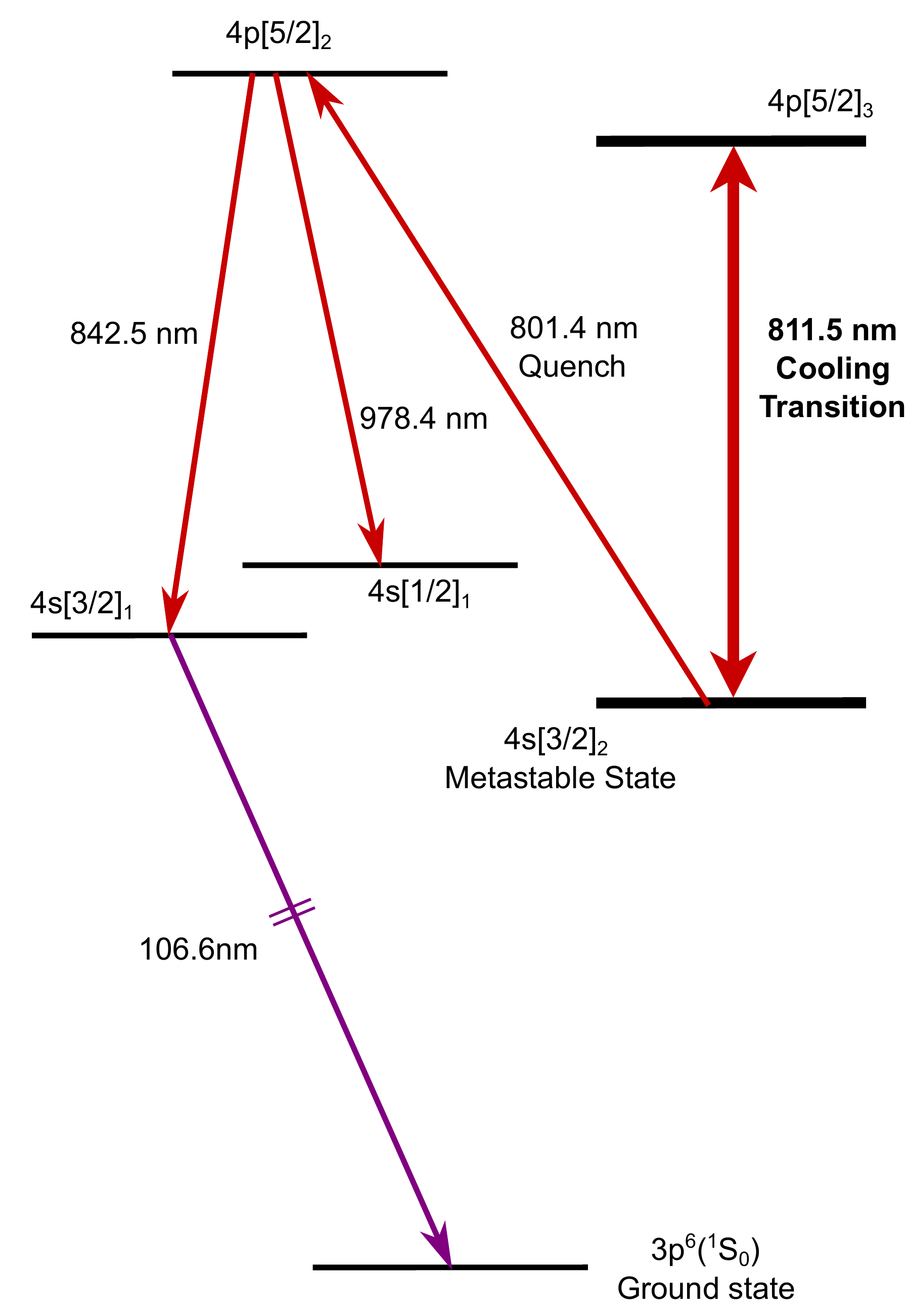}
\caption{Energy level diagram for argon, including relevant wavelengths for cooling and quenching to the ground state.}
\label{fig:energy}
\end{figure}

The trapping and detection of ground state argon atoms in a dipole trap is shown schematically in fig. 1.  Metastable argon atoms are first Doppler-cooled in a magneto-optical trap (MOT) on the 4s[3/2]$_2$ to 4p[5/2]$_3$ transitions \cite{katori1990laser}, as shown in fig. \ref{fig:energy}. A fraction of these atoms is then subsequently trapped in an optical lattice potential formed by an optical build-up cavity with a circulating intensity of $\sim$1kW that corresponds to a well depth of $\sim$2 mK for metastable argon atoms and $\sim$70$\mu$K for ground state atoms \cite{ref}. The trapped metastable atoms are transferred to the ground state by optically quenching from the 4s[3/2]$_2$ state using a laser operating at 801.4 nm. The quench laser first excites atoms from the 4s[3/2]$_2$ metastable state to the 4p[5/2]$_2$ state, from which they decay to the ground state via either the 4s[3/2]$_1$ or 4s[1/2]$_1$ states. The maximum photon recoil temperature from this process is 68 $\mu$K, which places a limit on the lowest temperature of the trapped ground state. By performing an incomplete quench we can populate the trap with both ground state and metastable species. 

The loading procedure for ground state atoms is similar to that which we used previously for metastable atoms \cite{ref}. The MOT is initially loaded for $\sim$2 seconds, after which time the intra-cavity intensity is ramped up over 30 ms to that required for trapping. After this period the quench beam is switched on for 1-5 ms (depending on how many atoms we want to quench) and the Zeeman slower beam, MOT magnetic field and MOT beams are all switched off and the atoms are held for the trapping period. To image atoms remaining after the trapping period, the MOT beams are switched back on and the remaining atoms are counted via their fluorescence recorded on a CCD camera. 

As metastable argon atoms have intrinsically high internal energy, an intra-trap collision can lead to either a Penning ($Ar^*+Ar^*\rightarrow Ar+Ar^++e^-$) or associative ionisation process ($Ar^*+Ar^*\rightarrow Ar_2^++e^-$) and in either process both metastable atoms are lost. If the trapped atoms are not spin-polarised, then for most metastable atoms the loss rate is of the order of 10$^{-10}$cm$^3$/s \cite{vassen} and trap lifetimes are ultimately limited by these interactions.

For trapping of metastable atoms only, trap loss from the optical lattice in the build up cavity can be described by a differential equation:

\begin{equation}
\dot{\rho_e}=-\Gamma\rho_e(t)-\gamma_{ee}\rho_{e}(t)^2
\label{diff} ,
\end{equation}

\noindent where $\rho_e$ is the density of trapped metastable atoms, $\Gamma$ is the one-body loss coefficient (i.e. mostly caused by collisions with background atoms) and $\gamma_{ee}$ is the two-body loss coefficient (caused by metastable intra-trap collisions). If the effective trap volume is not time-dependent, the number of trapped atoms is given by

\begin{equation}
\rho_{e}(t)=\frac{\Gamma\rho_e(0)e^{-\Gamma t}}{\gamma_{ee}\rho_e(0)(1-e^{-\Gamma t})+\Gamma}
 \label{eq:loss},
\end{equation}

\noindent where $\rho_e(0)$ is the initial density of metastable atoms.

The decay process when both metastable and ground-state atoms are trapped is more complicated. For this case we have two coupled differential equations, which describe the decay of both species:

\begin{equation}
\dot{\rho}_e=-\Gamma\rho_e-\gamma_{ge}\rho_g\rho_e-\gamma_{ee}\rho_e^2
\label{eq:coupled1}
\end{equation}

\begin{equation}
\dot{\rho}_g=-\Gamma\rho_g-\gamma_{ge}\rho_g\rho_e,
\label{eq:coupled2}
\end{equation}

\noindent where $\rho_g$ is the ground state density and $\gamma_{ge}$ is the loss coefficient during a ground state-metastable collision. To obtain an approximation to these equations we use the the Picard-Lindel{\"o}f theorem \cite{corduneanu2008principles,philip2002ordinary}, a description of which is given in the supplementary material.

Fig. \ref{fig:duallifetime} displays two lifetime curves; one loaded with only metastable atoms (squares) and the other with both ground state and metastable atoms (circles). Note that in each case the first few hundred milliseconds of the lifetime curve (when the density is highest) is dominated by Penning and associative losses, and the remaining loss after this time is primarily due to background collisions. To see this an additional line has been placed on the graph, demonstrating what the lifetime curve would be if only background collisions contributed. When both species are loaded the number of metastable atoms is observed to decrease more quickly when compared to the case of metastables only.

To verify this, equation \ref{eq:loss} was first fitted to the decay curve corresponding to metastable atoms only. Here the one-body loss coefficient, $\alpha$, is measured to be 1.3$\pm$0.1 s$^{-1}$ and the two-body loss coefficient, $\beta$, is  $(3.3\pm 0.8) \times 10^{-10}$ cm$^3$s$^{-1}$. Our two-body loss coefficient, $\beta$, is lower than a previously measured value \cite{PhysRevA.73.023406} of  $(5.8\pm 1.7) \times 10^{-10}$ cm$^3$s$^{-1}$. The previous value, however, is measured in a MOT without extrapolation to vanishing MOT light intensity. Light-assisted collisions will therefore artificially raise the measured value, which is in keeping with our lower value which is measured in the off-resonant lattice. 

Next, we fitted our approximate solution to equations \ref{eq:coupled1} and \ref{eq:coupled2} to obtain a value for $\gamma_{ge}$. To avoid over-parameterising the fitting routine, $\gamma_{ee}$ and $\Gamma$ were fixed at the already calculated values. In doing so, we found $\gamma_{ge}$ to be $(9\pm6)\times10^{-10}$cm$^3$s$^{-1}$.

\begin{figure}
\centering
\includegraphics[width=1.0\columnwidth]{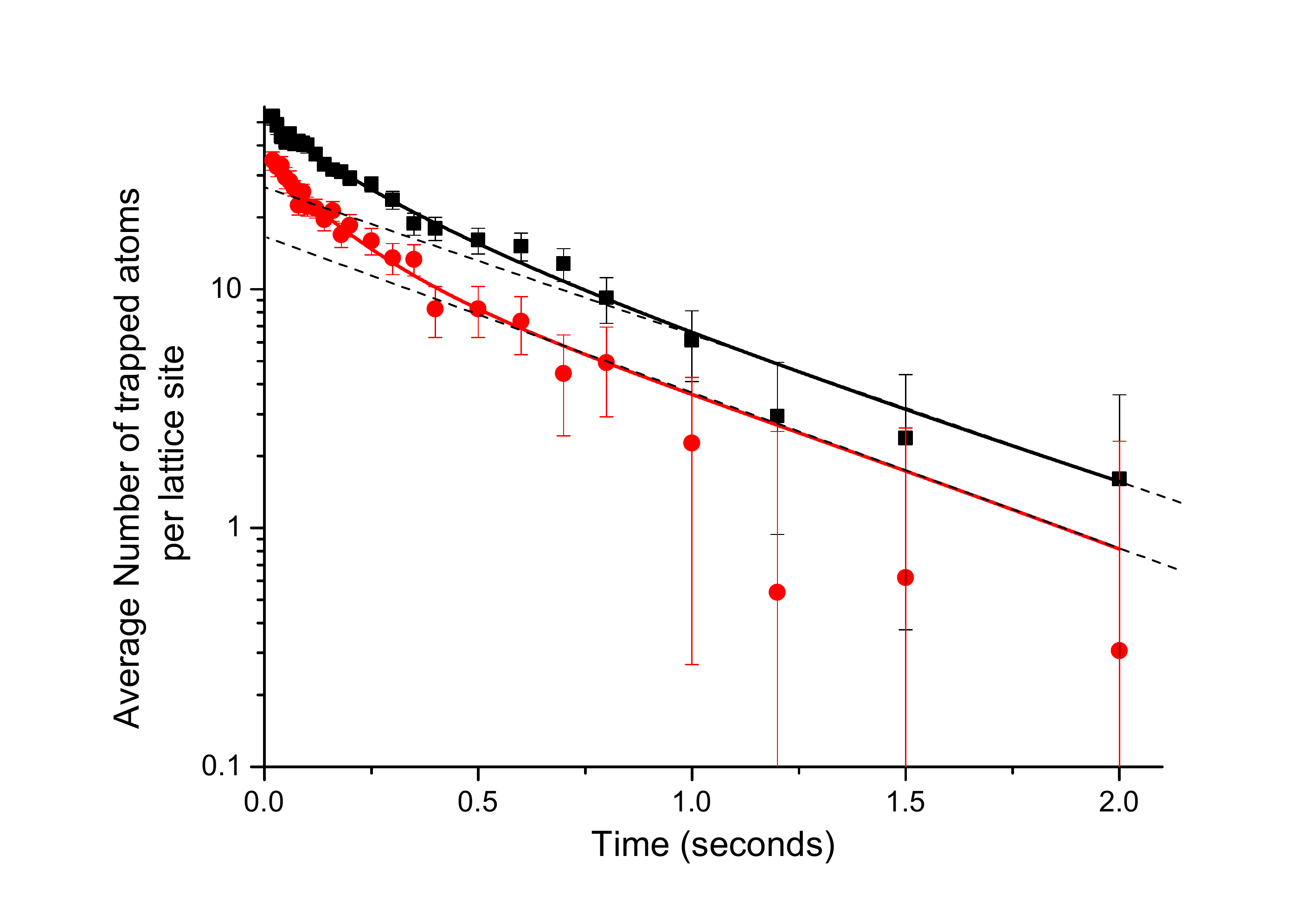}
\caption{Lifetime curves, displaying the average number of trapped metastable atoms in a the lattice trap as a function of time. The higher curve arises when only metastables are trapped. The lower curve is recorded when both metastable and ground state atoms are co-trapped. The dashed lines show how the atoms would decay if only collisions with background gases caused trap loss. These lines highlight how the early behaviour is dominated by Penning and associative losses.}
\label{fig:duallifetime}
\end{figure}

To verify that we have trapped ground state atoms we utilise a parametric heating technique. For a harmonic trap, the frequency of modulation at which significant heating and trap loss occurs is equal to $2 \omega/n$, where $\omega$ is the trap frequency and $n$ is an integer. The axial trap frequency in a harmonic trap is given by 

\begin{equation}
\omega_z=2\pi f_z=\sqrt{2\frac{U_0k^2}{m}},
\label{eq:freq}
\end{equation}

\noindent where $m$ is the mass of the trapped particle, $U_0$ is the well depth and $k=2\pi/\lambda$. $U_0$ is related to the polarizability by $U_0=\frac{2\alpha}{\epsilon _0 c} I_c$, where $I_c$ is the one-way circulating peak intensity. As the lattice wells are only harmonic for the lowest energy atoms the parametric heating spectrum is broadened. In addition, as we load both  ground and excited state atoms into the trap we expect to observe trap frequencies for both states. However as we only observe the metastable state, the loss for the ground state has a different signature from that of loss from the metastable state. This is because if ground state atoms are ejected from the trap the lifetime of the metastable atoms in the trap is increased since there are fewer collisions between the ground and metastable atoms. Instead of a decrease in observed fluorescence when modulated on a parametric heating resonance an increase is observed. When the trap frequency of the metastable atoms is reached, we observe the conventional parametric heating loss spectrum and a decrease in fluorescence. 

\begin{figure}
\centering
\includegraphics[width=1.0\columnwidth]{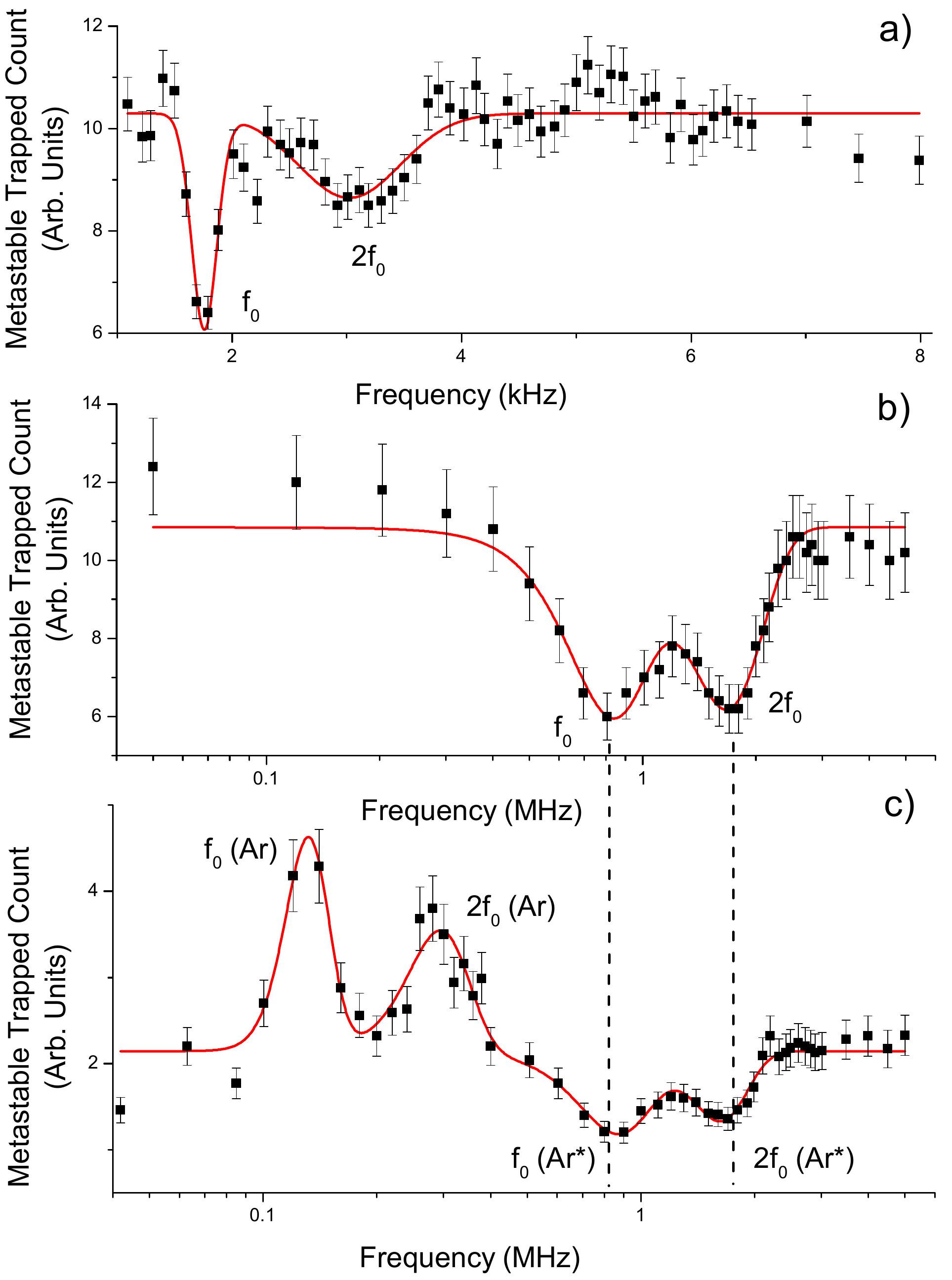}
\caption{Graph displaying parametric resonances of Ar and Ar* atoms observed in the dipole trap. a) shows two troughs due to metastable atoms being parametrically heated out of the trap at the radial trap frequencies. b) shows the same, but for the axial trap frequencies. c) shows the axial trap frequencies when we co-trap ground state and metastable atoms together. Here, the dips corresponding to the losses are the same case as in b). The peaks are due to ground state atoms being ejected out of the trap which reduces the collisional loss with metastable atoms.}
\label{fig:parametric}
\end{figure}

The trap frequencies for the metastable atoms were determined by applying a sinusoidal intensity modulation to the light coupled into the build-up cavity using an acousto-optic modulator. The well depth was modulated by 10$\%$ for frequencies up to 4 MHz for 100 ms. The trap was then turned off and the remaining atoms were imaged on an EMCCD camera following illumination by the MOT beams. This provided a parametric loss spectrum as shown in figure \ref{fig:parametric} a) and b).   

Fig. \ref{fig:parametric} a) shows two peaks corresponding to modulation at approximately the radial trap frequency and at twice this value, 1.7 kHz and 3.0 kHz respectively. Fig. \ref{fig:parametric} b) show two higher frequency peaks corresponding to modulation at the axial trap frequency and approximately twice this value along the lattice at 820 kHz and 1.67 MHz respectively.   These four frequencies correspond to a intracavity intensity of $\sim7\times10^9$ W/m$^2$ using a polarizability of 5.51$\times$10$^{-39}$ Cm$^2/$V \cite{PhysRevA.10.1131} in equation \ref{eq:freq}. To measure the ground state frequencies we quenched 80$\%$ of the metastables loaded into the trap. When the parametric loss measurements are repeated we observe four well-defined peaks as shown in figure c). Two of these peaks, at frequencies of 130 kHz and 300 kHz, correspond to the reduced loss of metastable atoms as ground state argon has been ejected from the trap. These frequencies correspond to the trap frequency and twice the trap frequency. The other two peaks at 890 kHz and 1.75 MHz show increased loss due to direct conventional parametric excitation of metastable argon out of the trap and are consistent with figure b). 

We use these trap frequencies to determine the ratio of the polarizability of metastable to ground state argon, $\alpha_{ar*}/\alpha_{ar}$. Using both parametric resonances at $f_z$ and $2f_z$ for both species, we calculate the $\alpha_{ar*}/\alpha_{ar}$ ratio to be 40$\pm$6. As the trap light is  far from any resonance the polarizability of the ground state should be well approximated by its static value given by 1.83$\times$10$^{-40}$ Cm$^2/$V \cite{mitroy2010theory}.  However, the metastable state polarizability is likely to be larger because it will be enhanced by the trapping light at 1064 nm.  We determine this polarisability by using the measured polarizability ratio and the assumption that the ground state value is well approximated by its static value. This gives a metastable polarizability of (7.3$\pm$1.1)$\times$10$^{-39}$ Cm$^2/$V, and as expected, this value is larger than the static polarizability given by 5.51$\times$10$^{-39}$ Cm$^2/$V.

In conclusion, we have trapped ground state argon atoms in an optical dipole trap and detected them using a parametric heating process. In contrast to typical parametric loss spectroscopy we detect the presence of ground state atoms in the trap by observing a reduced loss of co-trapped metastable argon. Using this method we have also measured the metastable to ground state polarizability ratio and from this the polarizability of the metastable state. By co-trapping both species we have also measured the one-body loss coefficient of metastable argon by the ground state argon and the two body loss coefficient for the metastable state in the absence of light assisted collisions. The measured one body loss co-efficient is 5 orders of magnitude larger than the value commonly used in models of argon plasmas \cite{bogaerts1995modeling}. As this is an important process in plasmas our value, determined at low temperature in these well controlled experiments, could be used to benchmark fully quantum collision calculations of this process.  Since our trap is designed for co-trapping molecules and atoms the parametric heating method will allow us to detect trapped molecules which, like ground state argon, often have transitions in the UV and VUV where laser sources are not readily available. In addition, by tuning slightly away from a parametric resonance we can in principle perform forced evaporation by selectively removing any hot trapped atoms or molecules \cite{poli2002cooling}. We note that although the metastable state density is always limited by Penning and associative losses the ground state is not. This may allow us in future to increase the density of ground state atoms well above 10$^{10}$cm$^{-3}$, which is important for sympathetic cooling using ground state argon or other laser cooled noble gas atoms.

\bibstyle{plain}

\bibliographystyle{apsrev4-1}

\end{document}